# A Systematic Review of Algorithmic Red Teaming Methodologies for Assurance and Security of AI Applications

Shruti Srivastava[1], Kiranmayee Janardhan[1], and Shaurya Jauhari[1, *]

[1] Responsible AI Office, Infosys Limited, India.

*Corresponding Author
Email: *shaurya.jauhari@infosys.com*

**Abstract.** Cybersecurity threats are becoming increasingly sophisticated, making traditional defense mechanisms and manual red teaming approaches insufficient for modern organizations. While red teaming has long been recognized as an effective method to identify vulnerabilities by simulating real-world attacks, its manual execution is resource-intensive, time-consuming, and lacks scalability for frequent assessments. These limitations have driven the evolution toward automated red teaming, which leverages artificial intelligence and automation to deliver efficient and adaptive security evaluations. This systematic review consolidates existing research on automated red teaming, examining its methodologies, tools, benefits, and limitations. The paper also highlights current trends, challenges, and research gaps, offering insights into future directions for improving automated red teaming as a critical component of proactive cybersecurity strategies. By synthesizing findings from diverse studies, this review aims to provide a comprehensive understanding of how automation enhances red teaming and strengthens organizational resilience against evolving cyber threats.

**Keywords:** Automation Testing, Generative AI, LLM Security, Red Teaming, Vulnerability Detection.

## 1    Introduction

Modern organizations operate in an interconnected digital ecosystem where cyber threats have grown highly sophisticated and persistent. Attackers now use advanced techniques like zero-day exploits, ransomware, and supply chain compromises, targeting vulnerabilities across networks and human factors. Traditional defenses struggle against adversaries leveraging automation and AI, while cloud computing, IoT, and remote work amplify complexity. Studies show cyber incidents are rising in frequency and severity, driven by technological advances and global interconnectivity[1,2]. These challenges highlight the need for innovative approaches like red teaming to simulate adversarial behavior and strengthen resilience.

To effectively counter the growing sophistication of cyber threats, organizations must adopt proactive strategies that go beyond traditional defensive measures. Red teaming has emerged as a critical approach in this context, offering a structured methodology to simulate real-world adversarial tactics and uncover hidden vulnerabilities before they can be exploited. Unlike traditional security checks, red teaming focuses on mimicking the mindset and techniques of attackers, enabling organizations to evaluate their resilience under realistic threat scenarios. This process not only identifies technical weaknesses but also exposes gaps in processes, policies, and human factors, thereby providing a holistic view of security posture[3]. Speaking of the context, while majority of the training and testing is confined to English language, there is a growing need to encompass other languages of geopolitical and economic significance, In a recent study, Cuevas et al. proposed a mechanism to red team probes in English, Spanish, and Hindi languages, for the supportive models[4]. This is extremely salient as organizations expand their geographic boundaries and scale businesses across global hotspots.

Generally, research demonstrates that red teaming significantly enhances an organization's ability to anticipate and mitigate risks by fostering a culture of continuous improvement and adaptive defense mechanisms. It changes security from reacting after problems happen to prevent them in advance, ensuring that organizations remain resilient in the face of emerging challenges. As cyber threats continue to evolve in complexity and scale, the importance of red teaming lies in its capacity to prepare organizations for worst-case scenarios, validate existing defenses, and strengthen overall security posture through realistic and rigorous testing. Furthermore, a threat is any potential source of harm, viz. a cyber-attacker, malware, system malfunction, or natural disaster, that could exploit a weakness. It represents what might cause damage but does not indicate whether it will occur. A risk, however, refers to the likelihood that a threat will successfully exploit a vulnerability and the severity of the impact if it does. In essence, threats are external or internal dangers that exist by themselves, whereas risks measure the probability and consequences of those dangers materializing. Effective risk management focuses not only on identifying threats but also on understanding vulnerabilities and implementing controls to reduce the chance and impact of harmful events[5,6].

While traditional red teaming is highly effective in identifying vulnerabilities, its manual execution presents significant challenges. Manual red teaming is resource-intensive, requiring skilled professionals, extensive planning, and considerable time to simulate realistic attack scenarios. This makes the process slow and costly, limiting its feasibility for frequent or large-scale security assessments. Additionally, manual approaches often lack scalability, as they cannot keep pace with the rapid evolution of cyber threats or the complexity of modern IT environments. Human-driven testing may also introduce inconsistencies and overlook certain attack vectors, reducing overall coverage. These limitations highlight the need for more automated, adaptive solutions that can deliver comprehensive and timely assessments without the constraints of manual operations.

The limitations of manual red teaming, such as high resource demands, slow execution, and lack of scalability, paved the way for automated solutions that leverage advanced technologies to enhance efficiency and coverage. Automated red teaming integrates artificial intelligence and machine learning to simulate adversarial behaviors at scale, enabling continuous and adaptive security assessments without the constraints of human-driven processes. This approach significantly reduces time and cost while improving consistency and comprehensiveness in identifying vulnerabilities. Automation also supports dynamic testing across complex environments, ensuring organizations can keep pace with evolving threats and rapidly changing infrastructures. Recent research emphasizes that effective automation not only accelerates red teaming but also augments human expertise, creating a hybrid model that combines machine precision with strategic human oversight[7,8]. As a standard, large language models (LLMs) are expected to follow "self-consistency", expectation that the outputs are near deterministic. We fathom determinism as a sampling-based reasoning strategy. This evolution marks a critical step toward proactive, scalable, and intelligent cybersecurity defense.

In this research paper, we will review how automated red teaming evolved as a solution to modern cybersecurity challenges. Since red teaming could also be performed manually, we also attempt an rejoinder to whether or not automated red teaming is necessary and sufficient condition to ensure AI security posture[9]. After discussing the rise of cyber threats, we will explore how automation and AI make red teaming faster, scalable, and more effective. The paper will cover current approaches, benefits, limitations, and future directions to understand how automated red teaming can strengthen security in today's dynamic threat landscape.

To ensure a rigorous and transparent review, we followed the PRISMA (Preferred Reporting Items for Systematic Reviews and Meta-Analyses) guidelines for identifying, screening, and selecting relevant studies[10]. The flow diagram (Figure 1) illustrates the complete process, including the number of records identified, screened, excluded, and finally included in this systematic review. Systematic reviews offer a mechanistic way to parse through erstwhile research literature and in this article we attempt to document the findings in both space and time[11].

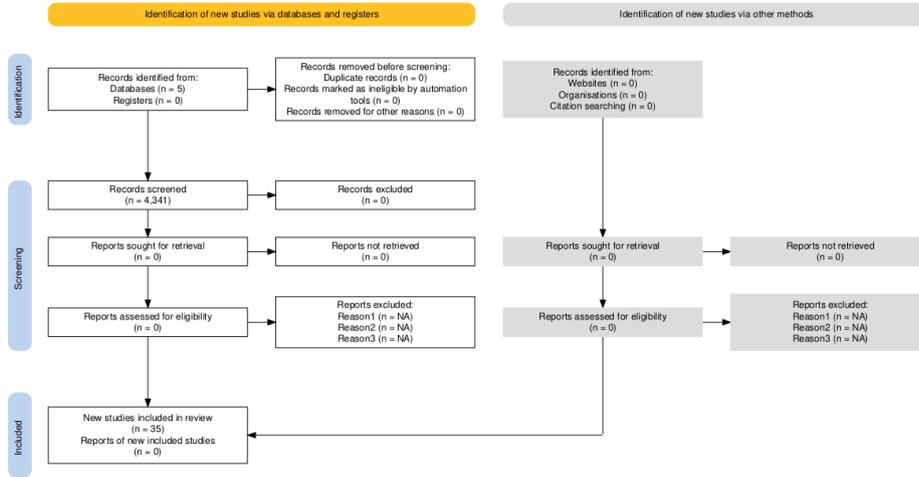

*Figure 1 PRISMA Flow Diagram depicting the systematic review on automated red teaming.*

## 2    Background and Definitions

Red teaming is an approach of flagging vulnerabilities in a target system. Specifically, from the context of Generative (Gen) AI systems, what Bender et al. would remark as "Stochastic Parrots"[12], given the non-deterministic and black-boxed behavior they exhibit, the implementation becomes immaculately articulate. The alignment of these systems of sticking to their programmed context of business domain, safety and societal obligations, ethics and legalities- are all expected to follow the realm of output. With the advancements in AI technologies enhancing their permeation, the nuanced output modes escalate the incumbency of red teaming exercises- text, image, audio, and video. We also understand that red teaming cannot be a one-time endeavor and needs to be cyclical in implementation, especially when the target application undergoes a structural revision.

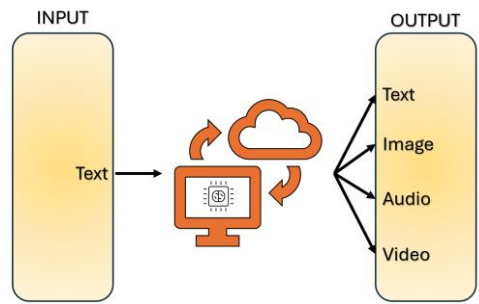

*Figure 2 For a typical text input given to a Gen AI model, a multi-modal output can be realized. That is not to discount that along with the text, any other supporting media can be supplemented as well.*

The Attack Success Rate (ASR) is a standard metric that exists across all corpus of studies and scientific literature, vis-à-vis evaluating the security posture of an AI application. In simple terms, the proportion of failed (jailbroken) responses in the total probes (inputs), optionally converted to a percentage, is an ASR. We highlight the following two observations though:

1.  ASR score is session-specific, and for auditibility purposes may not be reproducible. If a statistical analysis is to be performed, abiding by the law of large numbers, a sample of successive executions may render a ball park value that may be considered as a benchmark.

2.  Let us hypothesize that a banking AI chatbot is to be examined. Intuitively, the probes fed to the application must be domain specific and fall in the scope of banking and finance. If not, the ASR may reflect paradoxicality. We propose the negative and positive ASR scores in this context; a positive ASR establishing the security posture when the probes are contextually aligned, and a negative ASR when they are not, thus painting the complete picture through aggregation. Appropriate probing makes a significant difference, and hence comparing different applications (*apple to oranges*) on the basis of just a vague ASR may hit back adversely.

Let us ponder over a crucial backdrop of ASR. In terms of evaluation of a red teaming exercise, from the perspective of a binary classification problem, we are looking at two distinct response classes: *benign* and *jailbreak*. While the former represents that the response is acceptable, the latter highlights otherwise; effectively, these are our positive and negative classes, respectively. The classical performance (derived) metrics of the simulation, and not target application (model), can hence be determined, viz. precision, recall, accuracy, F1 score, etc.

|                       | Predicted: Jailbreak | Predicted: Benign   |
| --------------------- | -------------------- | ------------------- |
| **Actual: Jailbreak** | True Positive (TP)   | False Negative (FN) |
| **Actual: Benign**    | False Positive (FP)  | True Negative (TN)  |

In this confusion matrix, TP are cases where the system correctly identifies a positive event (for example, correctly flagging a real jailbreak attempt), while TN are cases where it correctly identifies a negative event (allowing a genuinely benign input). FP occur when the system incorrectly flags a negative case as positive, leading to unnecessary blocking or alerts, whereas FN occur when a positive case is missed and incorrectly treated as negative, allowing a real threat to pass through. Together, TP, TN, FP, and FN provide a complete picture of a system's detection performance, highlighting both its effectiveness at catching true issues and the trade-offs between security risk and over-restriction.

The derived metrics[1] provide a multidimensional view of system performance. Precision and specificity focus on avoiding unnecessary positives, recall and false negative rate emphasize catching all true positives, accuracy offers a broad overview, and the F1 score balances competing objectives. Selecting which metric to optimize depends on whether the priority is risk reduction, usability, or operational efficiency.

## 2.1    AI Applications

Building on the case for proactive, scalable red teaming, this section surveys AI application types from conventional ML and LLMs to model-embedded agentic systems, with emphasis on automated evaluation and operational guardrails. While classical ML (classification, anomaly detection, scoring) continues to support cybersecurity tasks, recent evidence shows a decisive shift toward LLM-centric application testing as organizations increasingly rely on large language models for both offensive and defensive workflows. Domain-specific, close-ended datasets such as CySecBench reveal harmful and policy-violating behaviors across intrusion, cryptographic, cloud, IoT, and web attack categories, demonstrating how LLMs can be systematically exercised using repeatable evaluation metrics such as success-rate and harmfulness scoring[13]. Moving beyond fixed corpora, ASTRAL shows that automated test generation using black-box coverage criteria, RAG-driven prompt synthesis, and LLM-as-oracle evaluation surfaces substantially more unsafe behaviors than static suites, illustrating why *dynamic*, automated approaches are critical for modern LLM-centric testing regimes[14].

Deployed AI systems depend heavily on structured measurement pipelines, where risk taxonomies, automated test generation, and evaluative critique model the safety requirements of real-world applications. S-Eval introduces a multi-dimensional safety taxonomy (eight dimensions, 100+ sub-risks) and an expansive bilingual benchmark for automated test generation and safety critique, enabling reproducible measurement across languages and attack types in operational contexts[15]. Complementary studies such as Adversarial Prompt Evaluation show that guardrail performance varies drastically by jailbreak style and that simple baselines may rival or exceed state-of-the-art detectors out-of-distribution, revealing dataset bias and the limitations of relying on any single detection model[16]. Further, Bypassing LLM Guardrails demonstrates that evasion strategies, e.g., invisible Unicode, homoglyphs, and adversarial ML perturbations, can achieve up to 100% bypass against widely deployed systems, reinforcing the requirement for layered measurement and multi-layered controls rather than static filters[17].

---

[1] See Definitions in Supplementary Data

As LLMs are embedded into real workflows, reading emails, browsing the web, generating code, orchestrating tools, organizations increasingly deploy agentic applications whose security depends on monitoring not just outputs, but internal reasoning and actions. LlamaFirewall operationalizes this insight through system-level guardrails: PromptGuard-2 (universal jailbreak detection), AlignmentCheck (goal-hijack and reasoning-path audit), and CodeShield (static analysis of generated code), providing defense-in-depth across the agent lifecycle[18]. At the base-model layer, the Security Assessment of DeepSeek and GPT Series underscores that architectures such as Mixture-of-Experts vs. dense RLHF-aligned transformers differ substantially in jailbreak robustness due to routing and alignment generalization dynamics, security-critical variables for agent deployments, not mere accuracy factors[19]. Similarly, **Security Steerability is All You Need** emphasizes measuring a model's ability to obey strict system-prompt guardrails under adversarial pressure, reframing evaluation from universal harms to task-specific operational constraints required for safe agentic applications[20].

Operational studies solidify the application landscape by examining real deployment realities. The study titled "*Lessons from Red Teaming 100 Generative AI Products*" identifies eight critical insights, automation increases coverage, human oversight remains indispensable, and red-teaming differs fundamentally from benchmarking, supporting continuous, multi-layered evaluation for agentic and LLM-centric systems[21]. Meanwhile, JailBench[2] demonstrates that multilingual, locale-specific applications require tailored benchmarks: its Chinese hierarchical taxonomy and jailbreak-enhanced dataset uncover deep safety vulnerabilities missed by English-centric datasets, underscoring the necessity of culturally and linguistically contextualized evaluation methods[22].

## 2.2 Datasets and Benchmarks

A central component of research on automated red teaming, LLM security, vulnerability detection, and agentic AI evaluation is the availability of standardized datasets and benchmarks that enable systematic, reproducible, and comparable assessment. Over the last four years (2022–2025), the field has seen an accelerated proliferation of such resources, responding to the rapid scaling of language models and the emerging need for rigorous security evaluation. Early foundational resources such as HELM and BIG-Bench provided broad coverage across capabilities, robustness, and safety dimensions, but subsequent work has produced increasingly specialized corpora targeting adversarial prompts, jailbreak behaviors, multilingual safety risks, toxic content generation, code-level vulnerabilities, and cyber-offense interaction patterns. In parallel, realistic environments and agent-task ecosystems, such as those simulating web navigation, API tool-calling, or enterprise workflows, have broadened the scope of evaluation beyond static prompts to dynamic, multi-step interactions. Collectively, these datasets and benchmarks capture the diversity of risk surfaces encountered by modern LLM-integrated systems, enabling researchers to probe structural weaknesses, measure the efficacy of guardrails, and test alignment generalization across contexts, languages, and threat models.

---

[2] Described later.

At the same time, this landscape reflects a clear methodological shift: from manually curated red-teaming prompts toward automated, large-scale, and domain-specific evaluation pipelines. Recent benchmarks incorporate richer taxonomies of safety risks, programmatically generated or adversely optimized test cases, and detailed metadata for traceability, severity scoring, and behavioral categorization. Several resources also bridge the gap between traditional software security and LLM-mediated vulnerabilities, introducing code-centric CVE-linked datasets, statement-level vulnerability corpora, and capture-the-flag challenge repositories. Others emphasize real-world operational realism, including community-sourced adversarial samples from public red-teaming events or enterprise-grade agent evaluations requiring planning, tool sequencing, and outcome verification. When integrated into a systematic review, these benchmarks provide a robust empirical foundation for mapping how automated red teaming has evolved, where current models fail, and which emerging risks remain insufficiently measured. The following table consolidates these datasets and benchmarks, offering the first comprehensive inventory spanning 2022–2025 across LLM security, generative AI evaluation, automation testing, red-teaming methodologies, and vulnerability-detection research.

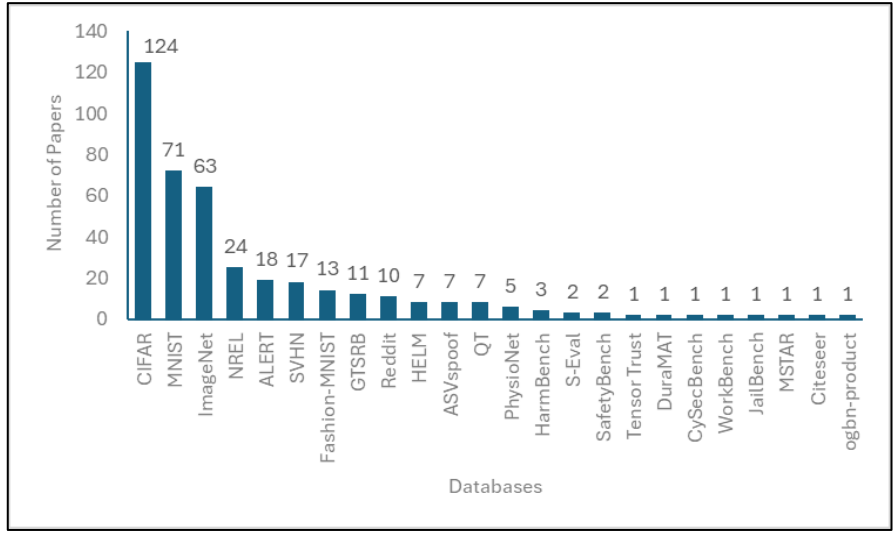

*Figure 3 While there is a slew of papers that generously highlight the image datasets, the benchmark datasets used in red teaming find sparse mentions in our distillation. Particularly, JailBench, HarmBench, CySecBench, SafetyBench, and S-Eval are the ones referred to in red teaming exercises.*

While not directly associated with red teaming, yet since they have emerged from our distillation of databases, let us cursorily acknowledge these important datasets before we reconcile with the datasets utilized in security benchmarking. The CIFAR[23], MNIST[24], and ImageNet[25] datasets fall into the broad paradigm of machine learning and artificial intelligence pertaining to image analysis, and have been widely referred in prediction studies (See Figure 3). These datasets have been sourced in umpteen image classification studies on object detection and segmentation. The MNIST dataset is one of the most well-known and foundational datasets in computer vision and machine learning. It consists of grayscale images of handwritten digits ranging from 0 to 9, with each image having a resolution of $28 \times 28$ pixels. The dataset contains 60,000 training images and 10,000 test images, making it relatively small and easy to work with. MNIST was designed to provide a standardized benchmark for evaluating classification algorithms, particularly neural networks employed for deep learning exercises. Due to its simplicity and low variability, it is widely used for educational purposes, rapid experimentation, and validating new model architectures. However, because these controlled images are highly constrained and lack real-world complexity, MNIST has not been able to cope up with the complexities of modern deep learning models, and hence lost viability as a challenging benchmark. This is where CIFAR and ImageNet datasets fare relatively well, though not without their own shortcomings.

The CIFAR datasets, developed by the Canadian Institute for Advanced Research, and that is where the acronym comes from, offer a more complex image classification challenge compared to MNIST. CIFAR-10 consists of 32 × 32-pixel RGB images across 10 object categories such as airplanes, automobiles, animals, and household objects, while CIFAR-100 extends this to 100 finer-grained classes. Both variants include 50,000 training images and 10,000 test images. Unlike MNIST, CIFAR images contain color information, background clutter, and greater intra-class variation, making them more representative of real-world visual tasks. As a result, CIFAR datasets are commonly used to benchmark convolutional neural networks, regularization techniques, and data augmentation strategies.

ImageNet is another large-scale, real-world image dataset created to advance research in visual recognition. It contains over 14 million labeled images spanning thousands of object categories, with the most used subset, ImageNet-1K, including around 1.2 million training images across 1,000 classes. Unlike MNIST and CIFAR, ImageNet images have variable and generally high resolutions, capturing significant diversity in lighting, viewpoints, backgrounds, and object scales. The dataset played a pivotal role in the deep learning revolution, particularly after the success of AlexNet in the 2012 ImageNet Large Scale Visual Recognition Challenge. ImageNet remains a standard benchmark for training and evaluating deep neural networks and is widely used for pretraining models that are later adapted to other vision tasks. Despite their impact, ImageNet suffers from high computational costs, potential bias, and diminishing benchmarking value, while CIFAR's low resolution limits its usefulness for detailed visual tasks. Several nuances of these datasets have been devised in recent years, as improvisation.

On red teaming per-se, our study distilled the occurrences of JailBench[26], HarmBench[27], CySecBench[28], SafetyBench[29], and S-Eval[15,30] datasets, that we briefly describe as below.

### 2.2.1 JailBench

JailBench is a Chinese-language jailbreak and security risk benchmark explicitly designed to expose deep, policy-evading vulnerabilities in LLMs operating in Mandarin and related linguistic contexts. Due to adoption of Chinese as one of the most spoken languages[31] and the government's push for research and development[32], Chinese is rivaling English as the preliminary language of Internet and knowledge hubs. We are likely to see increased adoption of the same, in the years to follow. Unlike earlier Chinese safety datasets that primarily focus on benign refusal scenarios, JailBench emphasizes active adversarial prompting, reflecting how real-world attackers attempt to bypass guardrails. It is aligned with China's official "Generative AI Service Security

Basic Requirements" standard, making it particularly relevant for regulatory and compliance-oriented evaluations in Chinese deployments.

The dataset is structured around a hierarchical safety taxonomy consisting of five high-level domains (e.g., harmful ideology, discrimination, illegal activities) and forty fine-grained subcategories. JailBench includes two interconnected subsets: *JailBench_seed*, which contains 540 base harmful prompts that models should reject, and *JailBench*, which expands these seeds into 10,800 highly optimized jailbreak prompts using an Automatic Jailbreak Prompt Engineer (AJPE). AJPE systematically mutates and evolves prompts using known jailbreak strategies, significantly increasing attack success rates compared to manually written adversarial inputs.

Empirical evaluations across 13 mainstream LLMs show that JailBench consistently achieves higher jailbreak success rates than prior Chinese benchmarks, including against commercial systems such as ChatGPT. These results underscore the benchmark's effectiveness at revealing latent vulnerabilities not surfaced by conventional refusal-based testing, particularly in culturally and linguistically nuanced contexts.

### 2.2.2 HarmBench

HarmBench is a standardized red-teaming benchmark and evaluation framework developed to address fragmentation and inconsistency in prior LLM safety assessments. While many earlier datasets combine heterogeneous prompt sources with ad-hoc metrics, HarmBench formalizes evaluation through a clearly specified threat model, behavior taxonomy, and reproducible scoring pipeline. Its primary focus is measuring a model's ability to robustly refuse harmful requests under adversarial pressure.

The dataset defines harmful behaviors along two axes: *functional categories* (standard, contextual, copyright, multimodal) and *semantic categories* (e.g., cybercrime, misinformation, chemical/biological harm, illegal activities). Contextual behaviors are particularly important, as they embed harmful intent within realistic narrative or personal contexts, which have been shown to substantially increase jailbreak success rates. HarmBench also releases a high-quality LLM-based behavior classifier, trained to approximate human judgments and achieving agreement levels comparable to GPT-4.

Beyond static evaluation, HarmBench is tightly coupled with automated red-teaming pipelines and adversarial training methods, such as Robust Refusal Dynamic Defense (R2D2). This design allows researchers to co-evolve attacks and defenses under controlled conditions, making HarmBench especially valuable for iterative safety training and longitudinal benchmarking of defense improvements.

### 2.2.3 CySecBench

CySecBench is a domain-specialized jailbreak benchmark targeting cybersecurity-related misuse of large language models. Its motivation stems from the observation that general-purpose harmful prompt datasets fail to adequately capture the technical structure and specificity of cybersecurity attacks, leading to noisy or overly broad evaluations. CySecBench addresses this gap by providing 12,662 close-ended prompts organized into ten well-defined attack categories, including malware development, cryptographic attacks, intrusion techniques, and web exploitation.

A distinguishing feature of CySecBench is its use of **close-ended prompts**, which constrain the expected malicious output space and allow for more consistent, quantitative measurement of jailbreak success. Prompts were generated and filtered using a structured pipeline to remove ambiguity and ensure technical validity. Experimental evaluations reveal that commercial black-box LLMs exhibit dramatically different success rates across cybersecurity subdomains, highlighting blind spots that are often obscured in general safety benchmarks.

CySecBench also introduces **prompt obfuscation–based jailbreak methods** and demonstrates that domain-specific adversarial strategies can outperform state-of-the-art general jailbreak techniques. As a result, CySecBench is particularly valuable for security-sensitive deployments, developer red teaming, and regulatory risk assessments involving cyber harm scenarios.

### 2.2.4 SafetyBench

SafetyBench departs from adversarial prompt benchmarking by focusing on safety knowledge and reasoning rather than generative jailbreak susceptibility. It consists of 11,435 multiple-choice questions spanning seven categories of safety concerns, including illegal behavior, ethics, misinformation, harassment, and personal harm. The dataset is bilingual (Chinese and English), enabling cross-lingual safety comparisons within a unified framework.

By using a multiple-choice format, SafetyBench reduces ambiguity in evaluation and allows for objective, reproducible scoring without reliance on subjective output classification. This makes it particularly useful for assessing a model's understanding of safety policies, rather than its ability to generate or refuse content under attack. Empirical studies unravel a strong gap between overall model capability and safety understanding, with even advanced models exhibiting inconsistent safety reasoning across categories.

SafetyBench is therefore best viewed as **complementary to jailbreak benchmarks**: it evaluates whether a model acknowledges unsafety, while adversarial datasets test whether it can consistently act on that knowledge when generating responses.

### 2.2.5 S-Eval

S-Eval represents a shift toward fully automated, large-scale safety evaluation of LLMs. It introduces a comprehensive risk taxonomy comprising **eight top-level** **dimensions and 102 fine-grained risks**, covering a much broader safety space than manually curated datasets. Using an LLM-based testing agent, S-Eval automatically generates over 220,000 evaluation prompts in both English and Chinese, eliminating the need for human annotation while dramatically increasing coverage.

The framework includes two specialized models: an *expert testing model* that generates adversarial test cases conditioned on risk categories, and a *safety critique model* that provides quantitative scores and explainable risk analyses. This design enables scalable, repeatable safety evaluation across rapidly evolving LLMs and attack strategies. Extensive validation shows that S-Eval identifies more diverse safety failures than traditional classifiers such as LLaMA-Guard-2, while remaining adaptable to new risk definitions. Because of its automation, scale, and adaptability, S-Eval is particularly well-suited for industrial deployment, continuous safety monitoring, and compliance-driven evaluations across multiple model versions and languages.

| Dataset | Primary Focus | Language(s) | Size | Evaluation Style | Key Strength | Typical Use Case | Web Reference(s) |
|---|---|---|---|---|---|---|---|
| **Jail-Bench** | Jailbreak vulnerability (Chinese context) | Chinese | 10,800 prompts | Adversarial prompting | Culturally aligned, high jailbreak success | Chinese-market safety audits, regulation-aligned testing | https://lin . .96-8186- |
| **Harm-Bench** | Automated red teaming and refusal robustness | English | ~500+ behaviors | Classifier-based adversarial evaluation | Standardized, reproducible red teaming | Defense benchmarking, adversarial training | https://arxiv.org/abs/2402.04249 |
| **CySec-Bench** | Cybersecurity | English | 12,662 prompts | Close-ended jailbreak testing | Domain-specific precision | Security-sensitive LLM deployment | https://arxiv.org/abs/2501.01335 |

| | | | | | | |
|---|---|---|---|---|---|---|
| | mis-use | | | | | |
| **Safe-tyBench** | Safety knowl edge and rea-soning | Chi-nese and Eng-lish | 11,43 5 MCQs | Multiple-choice QA | Objective, reproduci-ble scoring | Policy under-standing, cross-lingual safety analysis | https://acl anthol-ogy.org/2 024.acl-long.830/ |
| **S-Eval** | Com-pre-hen-sive risk dis-cov-ery | Chi-nese and Eng-lish | 220,0 00+ promp ts | Fully au-tomated LLM-based evalua-tion | Scale, cov-erage, adaptabil-ity | Continuous, large-scale safety monitoring | https://dl.a cm.org/do i/abs/10.1 145/37289 71, https://ui.a dsabs.har-vard.edu/a bs/2024ar Xiv24051 4191Y/ab-stract |

### 2.3 What is "algorithmic red teaming" and how does it differ from classic cyber red teaming?

In general, several closely associated terms are used in synonymy across domains. Emanating from the standard exercises of testing in software engineering and red team-ing in cybersecurity, red teaming of AI applications embraces the best of both worlds. A recent report from the Software Engineering Institute (SEI) finely demarcates this subtlety[33]. More technically, it explores how established practices in cybersecurity red-teaming can inform and strengthen Gen AI red-teaming. The authors have con-ducted parallel literature reviews of both domains to identify gaps and opportunities for improvement in AI safety testing and distill that Gen AI red-teaming is currently frag-mented and lacks standardized methodologies. **Most efforts focus narrowly on prompt-based jailbreaks and adversarial attacks, with limited threat modeling and inconsistent evaluation metrics**. In contrast, cyber red-teaming employs mature frameworks, including adversary emulation, multi-stage engagement processes, and structured reporting, which enable comprehensive vulnerability assessments.

To bridge this gap, the authors recommend formalizing threat modeling for Gen AI systems, adopting structured operational stages like those in cybersecurity, and improv-ing communication between red teams and system owners (and blue teamers) to ensure findings lead to actionable mitigations. They also emphasize the need for standardized manuals, community-driven best practices, and open-source tools to support repeatable[3] and auditable testing. Finally, the report calls for prioritizing known vulnerabilities such as prompt injection and jailbreaks, while developing systematic approaches for emerging risks like alignment to the timely recommendations made by standard organizations like OWASP[34], MITRE[35], NIST[36,37] and threat modeling directives by CSA[38]. By integrating these lessons, Gen AI red-teaming can evolve from ad-hoc prompt testing into a rigorous, end-to-end risk assessment process, which is essential as AI systems become increasingly embedded in critical applications.

Algorithmic (automated) red teaming typically involves three foundational components: an adversarial or attack model, a judge model, and a target model (the AI application under evaluation). Guided by an objective defined by the ethical tester, the attack model generates probing inputs designed to test the boundaries of the target model's safety mechanisms. These inputs are delivered to the target model, whose responses are then evaluated by the judge model. Based on its assessment, the judge determines whether a response is inappropriate, unsafe, or aligned with expected safety standards. An important nuance emerges in this workflow. As modern LLMs continue to evolve, their guardrails, safety layers, and policy alignment have become significantly more robust. Consequently, coaxing them into producing harmful, biased, or otherwise ill-founded prompts has become increasingly difficult. As a rule, older models, with less mature safety mechanisms, tend to be more susceptible to producing noxious or unsafe content, making them more effective as attack models. However, the opposite holds true for the judge model. Here, sophistication is critical: the judge must be nuanced, sensitive to context, and capable of accurately interpreting the sentiment, intent, and safety implications of generated responses. Hence, newer and more advanced models typically perform far better in this evaluative role, as they can render more precise, reliable judgments about the target model's behavior. On the same lines, the older attack models are preferred as they embrace least sophistication for generating probes; although this is not universally true.

## 2.4    Attack Classes

The decision boundaries of a target language model are inherently dynamic and can therefore be characterized as moving targets, continuously shifting as models evolve through updates, fine-tuning, and reinforcement learning processes[16,39]. Defensive mechanisms for Generative AI systems typically operate across multiple layers, including model-level alignment, input filtering, output moderation, and system-level guardrails. While reinforcement learning from human feedback (RLHF) and constitutional alignment have significantly improved baseline safety, empirical studies consistently show that no single defense mechanism is sufficient against adaptive adversaries[16,39]. Static keyword filters and rule-based classifiers, for example, are easily bypassed through paraphrasing, encoding tricks, or indirection, leading to a false sense of security when deployed in isolation[16,17,40]. A critical but often underexamined dimension of this dynamism concerns **implicit bias in guardrail design**.

---

[3] We argue that repeatability is fundamentally contrary to a Gen AI model's modus operandi. The same model may not necessarily engender the same ASR across multiple iterations, unless the premise is on factual accuracy.

Much like training-data bias, where a disproportionate share of pre-training and instruction-tuning data is drawn from English-language corpora, guardrail mechanisms are similarly optimized primarily for English inputs. As a result, safety filters, policy classifiers, and moderation heuristics may fail to accurately interpret or enforce constraints on probes crafted in non-English languages or in hybrid, code-switched expressions that combine multiple languages within a single instruction. This imbalance renders guardrails comparatively vulnerable to adversarial probes expressed in languages such as Japanese, Russian, German, or Italian, say, where linguistic structure, idiomatic usage, and cultural context differ significantly from English[4][26,41]. Such vulnerabilities are further compounded when adversaries introduce representational obfuscation techniques, including ASCII art, Base64 encoding, or other symbolic and encoded forms, that increase cognitive and computational burden on the detection layers. These transformations can mask intent while remaining interpretable to the generation layer, thereby selectively bypassing safety mechanisms, despite that the language models are trained on adversarial inputs too, with a view to expanding their wisdom on both negative and positive sentiments[17,40].

Although contemporary language models often possess translation capabilities that allow foreign-language inputs to be rendered into English prior to inference, this process is imperfect. Nuance, pragmatics, cultural connotations, and adversarial intent embedded in native expressions may survive translation or be partially lost, leading to incorrect safety judgments. Consequently, adversarial probes that exploit multilingualism and encoding strategies can traverse guardrails undetected, underscoring the need for multilingual, representation-aware, and context-sensitive defenses within algorithmic red-teaming frameworks[26,40,41]. Beyond text inputs, images, audio, and video manifest even further complications still. These modes can be tuned to embed harmful expressions that hoodwink the target AI applications. This circumstance represents even greater challenge, still, for the language models to defend[42-44].

Algorithmic red teaming plays a critical role in evaluating the real-world efficacy of defenses by stress-testing them under adversarial pressure. Automated probing reveals that defenses often perform well on in-distribution benchmarks but degrade sharply under novel jailbreak styles or out-of-distribution inputs. This phenomenon highlights Goodhart-style failures, where systems are optimized to pass known tests rather than to generalize robustly[16,30,39,45]. Consequently, defense efficacy must be measured longitudinally and comparatively, using diverse attack generators rather than static test-suites[30,45,46].

---

[4] We have alluded to the same fate befalling Mandarin too, yet due to the quantum of training data being more than other available languages, the inclusion gap with English is gradually narrowing.

A key insight from recent literature is that defense-in-depth architectures outperform monolithic approaches. Combining input-level detection, contextual policy evaluation, reasoning-path audits, and post-generation analysis yields more resilient systems, especially for agentic applications[18,47]. However, even layered defenses require continuous reevaluation, as improvements in alignment and filtering often coevolve with more sophisticated attack strategies. Thus, the efficacy of defenses should be viewed not as a fixed property, but as a moving equilibrium shaped by ongoing red-teaming cycles[16,18,30,39,45,46]. We would argue that this shortcoming of data bias would still extend to red teaming attempted by Agents as they would still call LLMs and tools that suffer from the same; so, scale is not a remedy[18,47].

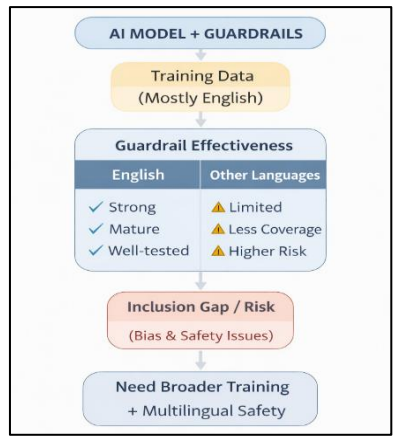

*Figure 4 A general observability is the innate bias to English language regarding training data, and the approaches following the responses from the AI Application.*

## 2.5    Efficacies of Defenses

Recent studies demonstrate that defense efficacy measured on static, single-turn suites substantially overestimate real-world robustness. Automated multi-turn red-teaming frameworks such as AutoAdv repeatedly rephrase, disguise, and refine an attack within a dialogue, achieving markedly higher ASR than single-turn baselines and revealing failure modes invisible to one-shot tests[46]. The core result is that alignment optimized for single exchanges often fails to persist across extended interactions, implying that realistic red-teaming must be multi-turn-aware and attacker-adaptive. A complementary, defense-centric critique shows that when attackers explicitly adapt to the design of the defense (gradient/RL/search/human-in-the-loop), they bypass a broad set of recent defenses with > 90% ASR, despite those defenses reporting near-zero ASR under static evaluations, evidence that robustness claims are fragile under adaptive threat models[38].

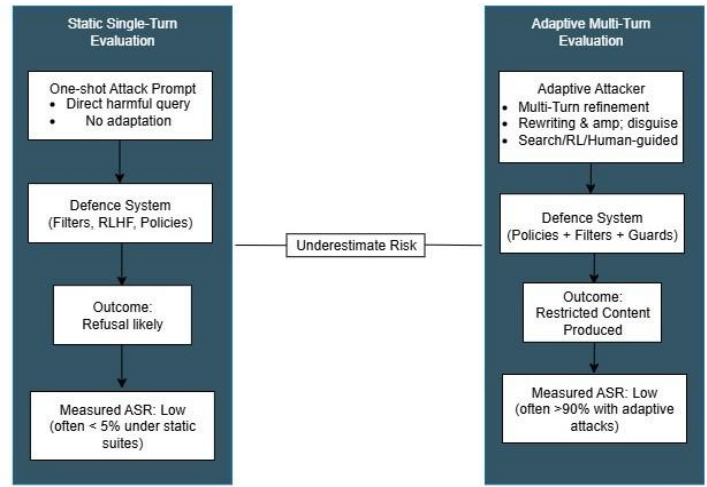

*Figure 5 Comparison of static single-turn versus adaptive multi-turn red-teaming evaluation for LLM defenses. Static evaluations rely on fixed, one-shot attack strings and often underestimate real-world vulnerability, producing artificially low ASR.*

Layered defenses help, but semantics and modality remain open doors. Benchmarking of camouflaged jailbreaks indicates that keyword and surface-pattern filters yield false assurance: malicious intent hidden in professional, mathematical, or engineering prose evades detectors while eliciting detailed, harmful instructions. The Behind the Mask benchmark (500 curated camouflaged prompts) operationalizes this threat and documents steep safety degradation under disguise, showing that semantics-level checks are required, not just lexical screening[40]. In the multimodal regime, MIRAGE demonstrates that narrative-driven visual-text sequences (role immersion, guided exploration) can lower defensive posture over turns and elicit restricted content from modern MLLMs; cross-modal cues and immersive context outperform text-only attacks and expose the brittleness of guardrails that are not modality-aware[43]. Together, these results suggest that defense-in-depth must include semantic normalization (e.g., paraphrase/decoding layers), context isolation, and modality-specific validators, all stress-tested under adaptive, multi-turn attack families rather than static, single-channel prompts.

Judge choice and meta-evaluation can invert conclusions about defense efficacy. A critical yet under-acknowledged factor is the vulnerability of LLM-as-judge evaluators that score jailbreak success and safety violations. Eiras et al. show that modest perturbations to output style and targeted adversarial manipulations **can shift a safety judge's false-negative rate by up to 0.24** on the same data, and in extreme cases, cause 100% of harmful generations to be misclassified as safe. Hence, the measured success of a defense can be an artefact of the chosen judge rather than the property of the defense itself. Robust practice therefore requires multi-evaluator protocols (diverse judges, rubrics, and human subsamples), disclosure of evaluator prompts, and sensitivity analysis to prevent evaluation-by-proxy from masking residual risk. When combined with adaptive attack frameworks, this meta-evaluation perspective reframes defense efficacy as a moving equilibrium that co-evolves with both attackers and evaluators, not a static benchmark score[48].

These findings align with the broader security-and-privacy survey literature that documents the persistent gap between curated evaluations and real-world misuse risk in LLMs/Multi-modal LLMs, reinforcing the need for attacker-adaptive, judge-robust, multi-modal, and multi-turn test harnesses for assurance claims. Surveys synthesize this trend and call for diversified, longitudinal measurement beyond single suites or single judges[49].

## 2.6 Progression of Automation from LLMs to Agents

LLMs are certainly efficient, if not effective, in complex red teaming exercises. Instead of rendering generic prompts, some approaches pool probes from the responses of the target model[50]. This scenario renders attack model the precise contextual information that upheaves plausibility of a jailbreak in consecutive sessions. This is remotely akin to the concept Agents adopt, with an innate ability to reason. Reasoning is what accentuates the role and capabilities of Agents, over LLMs, in application testing.

Automation from stand-alone LLMs to multimodal pipelines has expanded both adversarial coverage and realism. Early automated red teams formalized jailbreaks as optimization problems that learn from feedback, first in text-to-image (e.g., GhostPrompt, which couples dynamic prompt search with CLIP/safety-filter signals[51]), and then in text-to-video (e.g., T2V-OptJail, which frames jailbreaks as discrete prompt optimization[52]). These systems consistently bypass semantics-level filters while preserving intent, outperforming manual single-shot prompts. In parallel, GenBreak fine-tunes a red-team LLM via supervised signals and RL against a surrogate generator to find prompts that are both harmful and filter-evasive, strengthening the case for automated, learning-based red teaming in generative media. Together, these works establish a loop where the attacker derives new probes from model feedback, improving success and stealth across iterations[53].

Multimodal and context-aware attacks reveal qualitatively new failure modes—especially when semantics are split across channels. PiCo's "multimodal contextual jailbreaks" embed instructions inside code-styled images and exploit typographic cues and long-tail code priors to step through input filtering and runtime monitors, showing that cross-modal semantics can defeat keyword gates[44].Multimodal and context-aware attacks reveal qualitatively new failure modes, especially when semantics are split across channels. PiCo's "multimodal contextual jailbreaks" embed instructions inside code-styled images and exploit typographic cues and long-tail code priors to step through input filtering and runtime monitors, showing that cross-modal semantics can defeat keyword gates[44].

Complementing this, HIMRD (heuristic-induced multimodal risk distribution) spreads malicious intent across text and image, and uses a two-phase heuristic search to help the model *reconstruct* the latent request, achieving high black-box success rates against both open and closed MLLMs. These findings collectively argue that credible evaluation must be multi-turn and multimodal, since models can be induced, over several steps, to complete harmful goals that are never fully explicit in any single turn or modality.

As red teaming becomes stateful and tool-using, it converges on *agents* with planning, memory, and protocol-level reach. ARMs operationalize this shift: a plug-and-play adaptive agent that orchestrates 10+ multimodal strategies (e.g., reasoning hijacking, contextual cloaking), explores with ε-greedy search, and logs diverse failures at scale, reporting large uplifts in attack success and breadth across VLMs. Broader systematizations extend the threat surface from prompts to agent workflows, mapping input manipulation, model compromise, system/privacy attacks, and *protocol-layer* exploits in Model/Agent Context Protocol deployments, while LPCI formalizes persistent, delayed payloads hidden in agent memory/tool outputs that trigger across sessions. Meanwhile, *applied* studies like ScamAgents demonstrate agent-driven deception that adapts persuasively over turns, highlighting why assurance must evaluate full trajectories, not just isolated replies[54].

Defense and evaluation must coevolve with automation and agency. Trajectory-level benchmarks like SecReEvalBench introduce chain-aware metrics (resilience, refusal-logic, rejection-time) and multi-turn attack sequences (successive, reverse, ascending/descending severity) to quantify safety posture *over time*[55]. Surveys on LLM and agent security advise judge-robust scoring (ensembles, style-variance) and layered controls that go beyond I/O filtering to mechanism and protocol integrity[56]. In parallel, Moving Target Defense (MTD) for LLMs, ensemble routing and dynamic configurations, has shown large drops in jailbreak success, echoing mature MTD strategies from adjacent cyber domains and motivating agent-aware variants where models, tools, and policies rotate to raise attacker cost[47]. Finally, reinforcement learning's use on the *attacker* side (e.g., GAME-RL for malware example generation[57]) underscores the need for co-adaptive defenses that harden weights, validate memory/tools (against LPCI), and measure resilience under learning adversaries.

## 2.7 Compliance with Standards (NIST, OWASP, MITRE, EU AI Act, and ISO/IEC 42001)

Standards are not optional enablers but essential instruments for ensuring orderly coexistence, interoperability, and trust in a globally connected AI ecosystem. As AI literacy and adoption accelerate across sectors and jurisdictions, the absence of common baselines risks creating asymmetries in accountability, competitiveness, and regulatory assurance. To address this, established standards and frameworks define the minimum expectations for responsible development, deployment, and operation of AI systems. Conformance to these frameworks provides a defensible basis for determining compliance, demonstrating due diligence, and establishing the governance fitment of the AI application under evaluation.

### 2.7.1 NIST

The National Institute of Standards and Technology (NIST) is a U.S. federal agency under the Department of Commerce. It was founded in 1901 to promote innovation and industrial competitiveness by advancing measurement science, standards, and technology. In the context of AI adoption, what NIST defines first and foremost is a ***structure***: four verbs: **GOVERN**, **MAP**, **MEASURE**, and **MANAGE** that describe how an organization should run AI risk, beyond deliberation. The AI RMF 1.0 is voluntary, but it has quickly become the north star for creating auditable operating models around testing and assurance[58,59]. In practice, GOVERN sets the policies, roles, escalation paths, and accountability mechanisms under which red-team activities can happen safely (e.g., scoping approvals, fail-safes, incident handoffs, from the vantage point of the established RACI Matrix[5]). MAP forces us to articulate the context, stakeholders, impacts, data flows, system boundaries, so that red teaming is threat-informed rather than a grab-bag of jailbreak prompts. MEASURE is where the *rubber meets the road*: we instrument evaluation pipelines, define metrics (ASR, coverage, multilingual robustness, privacy extraction rate), and run adversarial tests continuously; this is a projection of the detection mechanism. Finally, MANAGE loops the results into risk treatment, remediation, and post-market monitoring, so issues do not reappear release after release. These expectations are codified in the AI RMF; they are intentionally out-come-oriented, letting us tailor our evidence, processes, and controls to our sector and risk posture[60].

---

[5] A RACI matrix is a project management tool used to clearly define roles and responsibilities for each task in a project. It outlines who is Responsible for doing the work, who is Accountable for final decisions, who needs to be Consulted for input, and who should be Informed about progress or outcomes. By mapping these roles in a simple table, a RACI matrix helps eliminate confusion, prevents overlaps or gaps in responsibilities, and ensures smooth coordination across teams.

Why does this matter specifically for GenAI? Because NIST added a Generative AI Profile in July 2024 that makes the abstract concrete for LLMs: it calls out prompt-in-jection risk, hallucination, leakage of sensitive data, and misuse at scale, and proposes actions we can transform into red-team objectives and acceptance criteria. A typical pattern we see is that teams align a red-teaming backlog to Profile items (e.g., indirect prompt injection via RAG sources; jailbreaks that escalate Agent permissions; privacy extraction via prefix-completion probing), then record runs, failures, mitigations, and regression results as evidence against MEASURE/MANAGE outcomes. The net effect is cultural: red teaming stops being "clever hacking" and becomes a governed measurement discipline with traceability from test to remediation to policy.

### 2.7.2 OWASP

If NIST is the operating model, OWASP is the field manual. Open Worldwide Application Security Project (OWASP) is a non-profit, global community focused on improving software and application security. It provides free, open-source resources, including tools, documentation, training, and widely adopted security standards such as the OWASP Top 10, which highlights the most critical web-application security risks. OWASP operates through chapters worldwide and supports open research, conferences, and community-led security projects. The Top 10 for LLM Applications began as an urgent response to production failures in 2023–24 and evolved into the broader GenAI Security Project, but its heartbeat is still pragmatic: enumerate the highest-leverage risks, provide concrete examples, and point to mitigations developers can deploy immediately. For red teams, this is gold[38].

Some defined vulnerabilities include "LLM01: Prompt Injection", that becomes a suite of adversarial prompts that include bilingual code-switching, Unicode confusable, zero-width characters, and indirect instructions planted in RAG documents or interaction probes. Likewise, "Insecure Output Handling" translates into tests that watch for XSS/CSRF/RCE when the model's output is fed to downstream components. "Training Data Poisoning" inspires backdoor trigger hunts and negative-prompt sweeps across fine-tunes. "Excessive Agency" and "Over-reliance" nudge us to simulate long-horizon agent tasks with unsafe tool selection and missing human-in-the-loop checks. Because the project is open and living, we can keep our test suites aligned to how the community sees systems fail, and not just how our internal team anticipates they might. These also cover flavors beyond LLMs to Agentic AI Security[61].

One practical way to use OWASP with LLM apps is to create **risk-to-tests mappings**. For each top-10 item, define seed prompts, mutation strategies (paraphrase, translation, obfuscation), policy-violation criteria, and remediation hooks. Automate this in CI/CD and collect artifacts per run, prompt, output, judgment, violation tag, to feed our governance and audit trails. This "attack library" is complementary to NIST's functions; OWASP tells us what to break, NIST tells us how to run and evidence the program.

### 2.7.3 MITRE

MITRE too is a not-for-profit organization that operates multiple Federally Funded Research and Development Centers (FFRDCs) for the U.S. government. It provides independent research, technical expertise, and guidance in areas such as cybersecurity, defense, AI, aviation, and healthcare. It plays a key role in advancing cybersecurity by enabling organizations to understand, detect, and defend against threats[35].

There are two MITRE worlds to keep straight. **ATT&CK** documents the tactics and techniques adversaries use to compromise enterprise systems, initial access, lateral movement, exfiltration, impact, and it remains essential whenever an AI system is embedded in a larger platform or supply chain. Red teams borrow ATT&CK to shape realistic kill-chains around the model; we can imagine compromised API keys that let an AI Agent call privileged tools, or action data exfiltration after a successful prompt-injection-induced code path.

Then there is **ATLAS**, which translates that adversary-informed discipline straight into the heart of AI/ML. ATLAS catalogs tactics and techniques specific to data, models, and Agent behavior. These include, yet are not limited to, poisoning training or retrieval corpora, adversarial examples and evasion, model extraction and inversion, prompt injection, and memory/goal manipulation in Agents. The content is grounded in real cases and red-team demos, which is why scenario design becomes easier; threats are emulated, not invented. A strong pattern is to chain ATLAS techniques (e.g., RAG poisoning to seed indirect prompts, jailbreak to escalate Agent actions, tool-misuse to hit a database) and then anchor the surrounding steps (credential abuse, persistence, exfiltration) in ATT&CK so our exercise covers both AI-specific and enterprise-security realities. This blend is especially effective for board-level drills and for engineering teams that must reconcile model safety with platform security.

### 2.7.4 EU AI Act

The EU AI Act is where "*should*" turns into "***must***". It entered into force on August 1, 2024[62,63]; obligations phase in over multiple years, prohibited practices and AI literacy by February 2, 2025, GPAI (foundation-model) obligations and governance chapters by August 2, 2025, and high-risk system requirements likely to land in 2026-2027, with further sandboxes and transparency duties coming online. Different actors (providers, deployers, importers, and distributors) carry different obligations, yet two ideas dominate:

- Risks[6] management must be systematic and documented, and
- There must be post-market monitoring tied to incident response.

For GPAI, expect transparency and data-governance duties; for high-risk systems, expect conformity assessment, technical documentation, and ongoing surveillance of real-world performance.

Where does red teaming fit? Right in the center of our technical file and post-market plan. We will need to show, for instance, that our model or Agent was tested against prompt-injection and jailbreak attempts, that we measured privacy/memorization leakage, that RAG pipelines were fuzzed for poisoning and indirect instructions, and that agentic workflows were exercised for unsafe tool use or goal hijack, especially in critical domains like healthcare[64]. Just as pertinent, we will need to prove that our findings led to mitigation, and that mitigations are regression-tested in subsequent product releases. **Without repeatable adversarial testing, it is hard to meet the Act's evidence expectations or withstand an audit by a national AI authority** as penalties, codes of practice, and oversight functions mature. Teams preparing now typically build a compliance pack per system: risk tiering rationale, threat models, red-team plans and runs, remediation records, monitoring results, and a playbook for incident classification and notification.

### 2.7.5 ISO/IEC 42001

The vein of ISO/IEC 27001 for information security is similar in rhythm to ISO/IEC 42001, i.e. a **Plan-Do-Check-Act management system** but applied to AI[65]. Published in December 18, 2023, it is the first international standard that tells an organization how to establish policy, roles, competence, lifecycle controls, supplier oversight, monitoring, internal audit, and continual improvement specifically for AI. It does not tell us how to build a model; rather educates on how to govern models in a way an auditor can verify. The key is to place our red-team program under *Operation* (scope, procedures, toolchains, safety gates), *Performance Evaluation* (metrics, internal audits, management review), and *Improvement* (nonconformity handling, corrective actions), we can transform adversarial testing into certification-grade evidence of Responsible AI. National adoptions (e.g., purported Australia's AS ISO/IEC 42001[7]) echo the same structure and emphasize lifecycle impact assessments and human oversight, making it straightforward to integrate with existing governance programs (27001, 27701, 9001). A subtle advantage of 42001 is that many enterprises already run ISO management systems, that can reuse mechanisms, document control, internal audit calendars, management reviews, supplier assessments, to make their AI red-team function sustainable rather than personality-driven.

---

[6] The different risk tiers defined in the EU AI Act range from Low to Critical.

[7] https://www.standards.org.au/news/standards-australia-adopts-the-international-standard-for-ai-management-system-as-iso-iec-42001-2023

Putting it all together, the most effective programs combine these lenses rather than choosing among them. MITRE ATLAS (with ATT&CK where relevant) can be used to craft threat-informed scenarios that reflect how adversaries target LLMs, RAG pipelines, and Agents. OWASP LLM Top-10 entries could be converted into concrete test families, prompt injection, insecure output handling, poisoning, excessive agency, and automate them with multilingual and obfuscation variants. Workflows and evidence under NIST AI RMF can be organized, so that red teaming becomes a continuous MEASURE/MANAGE loop informed by MAP and bounded by GOVERN. For the regulatory posture, the EU AI Act milestones can aid in aligning artifacts and monitoring, GPAI and high-risk obligations, and maintaining audit-ready technical files. Finally, whole practice can be embedded inside ISO/IEC 42001's AIMS so it survives staff changes, vendor shifts, and product pivots, and can be attested through independent audits.

## 3    Materials and Methods

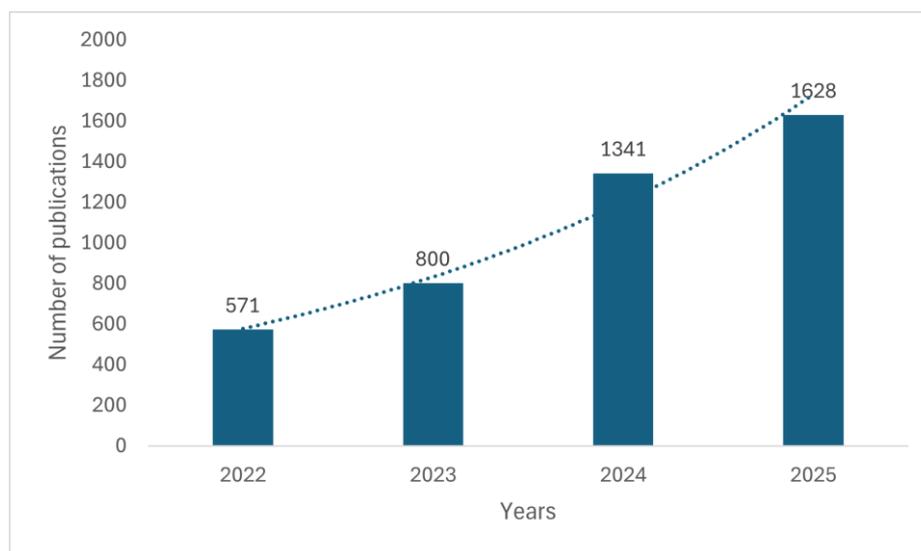

*Figure 6 Year-wise distribution of publications on automated red teaming (2022-2025); ~3× increase over four years.*

Above graph exhibits the rapid growth in the number of research publications on Automated Red Teaming, each year from 2022-2025, based on the consolidated dataset gathered for this systematic review. The yearly distribution shows a clear rise in research activity. Publications increased from 376 in 2022 to 569 in 2023, followed by a significant jump to 2,043 publications in 2024. Although the number decreases to 1,352 publications in 2025, the overall trend indicates a substantial growth in interest and work being conducted in this area.

This increase in publications reflects the expanding importance of Automated Red Teaming with the broader field of AI safety and evaluation. As AI systems become more capable and widely deployed, researchers are focusing more on methods that can automatically test vulnerabilities, assess model robustness, and support Responsible AI development. The spike in 2024 suggests heightened global attention toward building strong adversarial testing frameworks and improving automated evaluation techniques. These publications pattern reinforce the relevance of the systematic review, demonstrating that automated red teaming is a rapidly evolving research domain that requires a structured overview of existing methods, challenges, and opportunities.

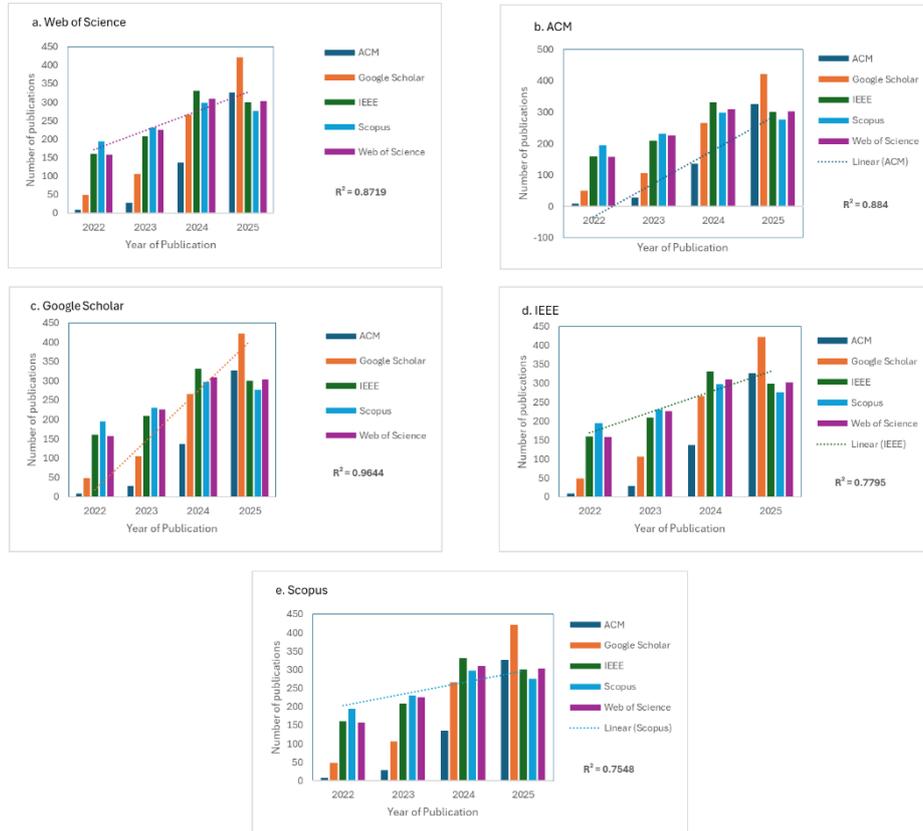

*Figure 7 Annual publication trends (2022–2025) across five databases with linear trendlines. Intuitively, we observe a positive trend in all databases that manifest the popularity of AI, particularly Gen AI.*

Annual publication counts from five major academic databases, viz. ACM Digital Library, Google Scholar, IEEE Explore, Scopus, and Web of Science were examined for the period 2022 to 2025 using a consistent and standardized search strategy. To analyze temporal trends, a linear regression model was applied separately to each database. The coefficient of determination ($R^2$) was calculated to assess how well the linear model explained changes in publication volume over time. Trendlines were plotted alongside bar charts to visually present both the observed publication counts and the fitted growth patterns.

The results indicate a clear and consistent upward trend in publication volume across all five databases throughout the study period. Among them, Google Scholar demonstrated the most stable and uniform growth, with an $R^2$ value of 0.9644, suggesting an almost perfect linear increase over time. ACM ($R^2 = 0.08840$) and Web of Science ($R^2 = 0.8719$) also exhibit strong linear relationships, reflecting steady and predictable year-on-year growth in indexed publications.

In contrast, IEEE ($R^2 = 0.7795$) and Scopus ($R^2 = 0.7548$) showed growth trends with comparatively greater variability. While both databases experienced overall increases in publication counts, the lower $R^2$ value suggests moderate fluctuations, potentially attributable to variations in annual research output, indexing practices, or databases' coverage policies.

Overall, the trendline analysis confirms a sustained rise in research activity within the domain examined from 2022 to 2025. The consistently high $R^2$ values observed across most databases indicate that this growth is systematic rather than sporadic. Google Scholar reflects the most uniform expansion, underscoring the growing scholarly interest in the topic and its broad dissemination across multiple academic indexing platforms.

To perform this systematic review, we considered the following search terms, across five databases, as aforementioned. As much as we would have preferred, we were not able to utilize data from Arxiv (https://arxiv.org/), which is a double-edged sword. While most of the articles will be hosted [there], we can conclude that these are not peer-reviewed. Subsequently, they will find way into journals/conferences, and even that, in relatively better shape.

***"automated", "generative adversarial examples (GAEs)", "algorithmic", "adversarial attacks", "vulnerability testing", "jailbreak", "AI Security", "security benchmarking", "red teaming framework", "LLM security", "prompt injection", "hate speech detection", "privacy violation", "evasion attacks", "poisoning attacks", "black-box attacks", "harmful behavior", "profanity", "stress testing AI"***

The search terms were fed to the search operation in the formatting expected by the respective databases. For instance, the following is the complete string from Scopus.

*ABS ( {automated} OR {generative adversarial examples (GAEs)} OR {algorithmic} OR {adversarial attacks} OR {vulnerability testing} OR {jailbreak} OR {AI Security} OR {security benchmarking} OR {red teaming framework} OR {LLM security} OR {prompt injection} OR {hate speech detection} OR {privacy violation} OR {evasion attacks} OR {poisoning attacks} OR {black-box attacks} OR {harmful behavior} OR {profanity} OR {stress testing AI} ) AND PUBYEAR > 2021 AND PUBYEAR < 2026 AND ( LIMIT-TO ( OA , "all" ) ) AND ( LIMIT-TO ( SUBJAREA , "COMP" ) ) AND ( LIMIT-TO ( LANGUAGE , "English" ) )*

As is notable in the above example, the adopted inclusion criteria, for searching databases in general are included in the staple conditions below.

1. Time Range: **January 2022 to December 2025**
2. Subject Area: **Computer Science**
3. Language Medium: **English**
4. Access status: **Open Access** only
5. Searched Within (wherever provisioned): **Abstract**

| S. No. | Database | Accessed Weblink | Total records gleaned | Top records considered after filtering |
|--------|----------|------------------|-----------------------|----------------------------------------|
| 1 | ACM Digital Library | https://dl.acm.org/ | 41460 | 500 |
| 2 | IEEE Explore | https://ieeexplore.ieee.org/Xplore/home.jsp | 3500 | 1000 |
| 3 | Web of Science | https://clarivate.com/academia-government/scientific-and- | 5690 | 997 |

| | | academic-re-search/re-search-dis-covery-and-referenc-ing/web-of-science/ | | |
|---|---|---|---|---|
| 4 | Google Scholar | https://scholar.google.com/ | 1250000 | 843 |
| 5 | Scopus | https://www.sco-pus.com/pages/home#basic | 35396 | 1000 |

## 4      Discussion and Future Directions

The rapid evolution of automated red teaming reflects a broader shift in how organizations evaluate and secure AI systems. As our review indicates, automated adversarial testing frameworks, spanning text-only jailbreaks, multimodal prompting, and agent-based exploration, have significantly expanded the breadth and repeatability of red-team assessments. Furthermore, alignment to the standard frameworks has seen wide adoption, arguably due to the enterprise compliance obligations (See **Tabulating Occurrences** in the **Supplementary Data** section). Yet, despite this progress, several structural and methodological challenges reveal that the field is still in an early stage of maturity. Current approaches are often optimized for narrow, benchmark-driven evaluations rather than realistic threat models. Many defenses that appear robust under static test suites degrade sharply when subjected to adaptive or multi-turn attacks, demonstrating a Goodhart-like tendency for systems to "***pass the test but fail the goal***". This indicates that red-teaming effectiveness must be judged not by isolated metrics but by how systems perform under evolving, context-aware, and often ambiguous adversarial pressures.

Numerous studies discuss guardrail frameworks and moderation models, yet many demonstrate high evasion/jailbreak success rates against popular defenses, reinforcing the need for defense-in-depth and continuous, automated testing. The guardrails exist, but remain bypassable, leaving the broader system vulnerable to the AI Application that apparently represents the "start-point" for a bad actor. Hence, derisking the AI systems is of paramount importance for the uprightness of entire infrastructure. In terms of models and data, safeguarding the proprietary information, and ensuring trust and maintaining output within the realm of enterprise safety for the input/ interface layer.

A recurring theme across literature is the fragility of guardrails when confronted with multilingual, culturally contextual, or multimodal probes. Most safety layers, classifiers, and moderation systems are trained predominantly on English language data, resulting in inconsistent performance for other languages where idioms, pragmatics, and cultural norms differ substantially. Benchmarks such as JailBench high-light how safety mechanisms that seem effective in English can be trivially bypassed in Mandarin or other languages. This linguistic asymmetry becomes even more critical as global organizations deploy AI systems across multi-lingual user bases. Simi-lar vulnerabilities emerge in multimodal systems, where text-to-image and multi-turn visual-text interactions create additional pathways for bypassing safety filters. Newer studies show that harmful intent can be distributed across modalities, latent in images, subtly encoded in visual representations, or revealed only through cross-modal reasoning, rendering traditional text-centric defenses insufficient.

Another major challenge lies in the reliability of automated judging. LLM-as-judge frameworks have become standard due to their scalability, yet empirical evidence demonstrates that even minor perturbations in output style, formatting, or reasoning patterns can significantly affect their safety assessments. This variability means that measured attack success rates may reflect the weaknesses of the judge rather than the strengths of the defense or vulnerabilities of the target model. As a result, safety evaluation must move toward more robust judge ensembles, cross-model consensus mechanisms, and periodic human-in-the-loop validation, especially for high-risk categories such as cybersecurity misuse, harmful biological content, and autonomous agent behavior.

The emergence of AI Agents introduces an entirely new attack surface and necessitates a rethinking of red-teaming objectives. Unlike conventional LLMs, agents operate across multi-step tasks, maintain internal memory states, and can invoke ex-ternal tools and APIs. This makes them susceptible not only to jailbreaks but also to reasoning-path manipulation, tool-use hijacking, goal hijack, long-term memory poisoning, and latent payloads that trigger across sessions. Red-teaming such systems requires trajectory-level evaluation frameworks that assess behavior over the full lifecycle of interactions rather than isolated responses. Early agent-focused red-team tools demonstrate that even well-aligned models can drift into unsafe actions under prolonged or strategically structured prompting, indicating that persistence, context accumulation, and emergent behavior must become central areas of evaluation.

Looking ahead, the future of automated red teaming must focus on several transformational directions. First, multilingual and culturally grounded adversarial testing must become foundational rather than peripheral, ensuring that evaluations reflect the global deployment landscape rather than a narrow linguistic subset. Second, multimodal red teaming should expand substantially, with new datasets, attack families, and evaluation methodologies that stress-test the complex cross-modal reasoning abilities of modern models. Third, judge robustness must be strengthened through ensemble methods, perturbation-aware scoring, and transparency in evaluation prompts and criteria. Fourth, agentic AI requires the development of novel threat models, memory-aware attacks, and chain-of-thought-sensitive evaluation tools that examine safety at the level of plans, sub-goals, and emergent trajectories.

Finally, and critically, automated red teaming must integrate more deeply with governance and compliance frameworks. Standards such as the NIST AI RMF, OWASP LLM Top 10, MITRE ATLAS, EU AI Act, and ISO/IEC 42001 call for continuous, documented, and risk-informed testing. As organizations move toward regulated AI operations, automated red teaming will need to produce audit-ready artifacts, sup-port post-market monitoring, and demonstrate traceable links between findings, mitigations, and governance outcomes. The field is therefore shifting from clever adversarial prompting toward disciplined, lifecycle-integrated assurance engineering.

## 5     Conclusion

We understand that automated red teaming is no longer optional, albeit foundational to the security, reliability, and trustworthiness of modern AI systems. The rapid evolution of LLMs, multimodal models, and agentic architecture demands adversarial testing methodologies that are continuous, scalable, and adaptive. Through analysis of state-of-the-art frameworks, datasets, and defense mechanisms, we find that automated red teaming consistently uncovers vulnerabilities invisible to manual or static evaluations, particularly in multilingual, multi-turn, and multimodal contexts. A practical red-team program should rely primarily on automated, tool-driven testing to continuously uncover model and system vulnerabilities, while reserving hu-man experts for the most complex or novel attack scenarios. Effective evaluation must consider the entire deployed system, including RAG components, plugins, agents, and infrastructure, and extend across all supported modalities. To ensure broad and adaptive coverage, teams should combine diverse, established safety benchmarks with domain-specific and dynamically generated tests. Finally, guard-rails should be treated as just one layer within a defense-in-depth strategy that also includes policy-steering, runtime monitoring, retrieval sanitization, agent oversight, and secure fine-tuning practices to prevent safety regressions.

At the same time, the review highlights key limitations in current practice, including evaluator fragility, limited domain and linguistic scope, and the challenges of modeling realistic, agentic adversaries. Standards and governance frameworks are beginning to formalize expectations for AI security testing, signaling a shift toward more recognized, rigorous and accountable assurance practices.

Overall, the evidence suggests that the future of AI safety lies in hybrid approaches: scalable automation augmented with human oversight, multi-layered defenses tested by multi-modal attacks, and red-teaming pipelines tightly integrated with organizational governance and regulatory compliance. As models grow more capable and pervasive, automated red teaming will remain essential to ensuring that AI technologies are resilient, responsible, and aligned with societal values.

In conclusion, automated red teaming is on the cusp of becoming a pillar of AI safety, but its methodologies must keep pace with the complexity, autonomy, and ubiquity of modern AI systems. Future research should emphasize adaptive, multilingual, multimodal, agent-centric, and judge-robust approaches, moving from static evaluation toward dynamic, realistic, and continuously evolving adversarial assessment. Doing so will ensure that red teaming not only identifies vulnerabilities but also strengthens resilience, trustworthiness, and regulatory readiness across the entire AI ecosystem.

**Conflict of Interest**

None Declared

**Acknowledgement**

The authors gratefully acknowledge the steadfast support of the Responsible AI Office, Infosys Limited. Also, Microsoft Copilot was utilized as a tool for improving the clarity and readability of the manuscript draft. However, the tool did not contribute to the analysis, interpretation, or generation of scientific content.

## Appendix/ Supplementary Data

**Definitions**

- **Precision** measures how reliable positive predictions are. It answers the question: *of all cases predicted as positive, how many were actually positive?*

$$\text{Precision} = \frac{TP}{TP + FP}$$

- **Recall** (also called **Sensitivity** or **True Positive Rate**) measures how effectively the system captures actual positive cases. It answers: *of all real positives, how many did the system detect?*

$$\text{Recall} = \frac{TP}{TP + FN}$$

- **False Positive Rate (FPR)** ascertains how often negative cases are incorrectly classified as positive.

$$\text{FPR} = \frac{FP}{FP + TN}$$

- **False Negative Rate (FNR)** reflects on how often positive cases are incorrectly classified as negative.

$$\text{FNR} = \frac{FN}{FN + TP}$$

- **Accuracy** measures overall correctness across all predictions, though could still represent bias for either negative or positive classes.

$$\text{Accuracy} = \frac{TP + TN}{TP + TN + FP + FN}$$

- **F1 Score** is the harmonic mean of precision and recall, used to balance the trade-off between the two.

$$F1 = 2 \times \frac{\text{Precision} \times \text{Recall}}{\text{Precision} + \text{Recall}}$$

- **Specificity** (also called **True Negative Rate**) gauges how well the system identifies negative cases correctly.

$$\text{Specificity} = \frac{TN}{TN + FP}$$

**Tabulating Occurrences**

In the derived data, we further carried out lexical search for specific terms like "***algorithmic***", "***standards***", "***frameworks***", "***RAG***", "***tools***", "***Multi-agent systems***", "***roleplay***", "***prompt injection***", "***jailbreak***", amongst others, in the titles, keywords and abstracts of the papers. The associated counts are reflected below. Even though the broad research primarily emphasized 'algorithmic red teaming' and 'automation' implicitly accounted for many of the hits, the similarly high occurrence of 'taxonomy' and 'frameworks' signals the continued centrality of governance and standardization frameworks.

| S. No. | Approaches | Count |
|---|---|---|
| 1 | Evaluation/ Taxonomy/ Frameworks | 1264 |
| 2 | Automated/ Programmatic | 1224 |
| 3 | Optimization/ Black-Box Attacks | 829 |
| 4 | RAG/ Tool/ Agent Integration | 663 |
| 5 | Multimodal | 522 |
| 6 | Safety Benchmarks/ Datasets | 346 |
| 7 | Guardrails/ Moderation | 302 |
| 8 | Multi-agent | 286 |
| 9 | Prompt obfuscation/ roleplay/ persona | 25 |